# Status of the SIMP project: Towards the Single Microwave Photon Detection

**David Alesini [1] • Danilo Babusci [1] • Carlo Barone [2,3] • Bruno Buonomo [1] • Matteo Mario Beretta [1] • Lorenzo Bianchini [4] • Gabriella Castellano [5] • Fabio Chiarello [5,1] • Daniele Di Gioacchino [1] • Paolo Falferi [6,7] • Giulietto Felici [1] • Giovanni Filatrella [8,3] • Luca Gennaro Foggetta [1] • Alessandro Gallo [1] • Claudio Gatti [1] • Francesco Giazotto [9,4] • Gianluca Lamanna [10,4] • Franco Ligabue [4] • Nadia Ligato [4] • Carlo Ligi [1] • Giovanni Maccarrone [1] • Benno Margesin [11,7] • Francesco Mattioli [5,1] • Eugenio Monticone [12,13] • Luca Oberto [12,7] • Sergio Pagano [2,3] • Federico Paolucci [4] • Mauro Rajteri [12,13] • Alessio Rettaroli [14,1] • Luigi Rolandi [4] • Paolo Spagnolo [4] • Alessandra Toncelli [4] • Guido Torrioli [5,1]**

*1) INFN, Laboratori Nazionali di Frascati, Frascati, Roma, Italy*
*2) Dipartimento di Fisica, Università di Salerno, Salerno, Italy*
*3) INFN, Gruppo Collegato di Salerno, Salerno, Italy*
*4) INFN, Sezione di Pisa, Pisa, Italy*
*5) Istituto di Fotonica e Nanotecnologie CNR, Roma, Italy*
*6) Istituto di Fotonica e Nanotecnologie CNR - Fondazione Bruno Kessler, Povo, Trento, Italy*
*7) INFN, TIFPA, Povo, Trento, Italy*
*8) Dipartimento di Scienze e Tecnologie, Università del Sannio, Salerno, Italy*
*9) NEST, Pisa, Italy*
*10) Dipartimento di Fisica, Università di Pisa, Pisa, Italy*
*11) Fondazione Bruno Kessler, Povo, Trento, Italy*
*12) Istituto Nazionale di Ricerca Metrologica (INRIM), Torino, Italy*
*13) INFN, Sezione di Genova, Genova, Italy*
*14) Dipartimento di Fisica, Università di Roma Tre, Roma, Italy*

**Abstract** The Italian institute for nuclear physics (INFN) has financed the SIMP project (2019-2021) in order to strengthen its skills and technologies in the field of meV detectors with the ultimate aim of developing a single microwave photon detector. This goal will be pursued by improving the sensitivity and the dark count rate of two types of photodetectors: current

Author 1 • Author 2 • Author 3

biased Josephson Junction (JJ) for the frequency range 10-50 GHz and Transition Edge Sensor (TES) for the frequency range 30-100 GHz. Preliminary results on materials and devices characterization are presented.

**Keywords** Transition Edge Sensor • Josephson Junction • SQUID • Axion

## 1 Introduction

The low-mass frontier of Dark Matter, the measurement of the neutrino mass, the search for new light bosons in laboratory experiments, all require detectors sensitive to excitations of meV or smaller. Faint and rare signals, such as those produced by vacuum photoemission or by an Axion in a magnetic field as in the Quax experiment [1], could be efficiently detected only by a new class of sensors.
For this purpose, INFN has financed the SIMP project which has the goal of developing two types of single microwave photon detectors: Transition Edge Sensor (TES) for the frequency range 30-100 GHz and current biased Josephson Junction (JJ) for the frequency range 10-50 GHz.

## 2 Transition Edge Sensor

Transition Edge Sensors are microcalorimeters based on the steep resistive transition of a superconducting material [2]: superconducting films are biased within the transition region [Fig. 1] where the high slope of the resistance vs temperature curve enhances the sensitivity to temperature variations and makes them thermometers suitable either as bolometers or as microcalorimeters.

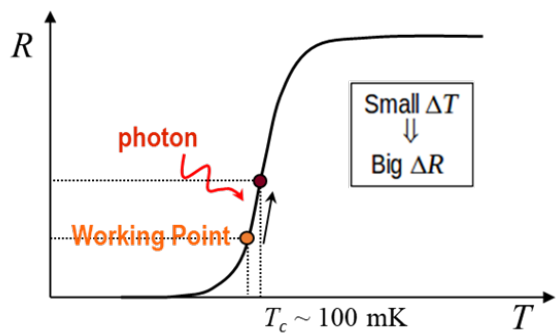

**Fig. 1** Working principle of a single-photon TES. Thanks to a sharp superconducting transition small changes in temperature cause large changes in resistance.

# Title

While the first demonstration of a TES dates back to the 40s, the read out technology became reliable only in the 90s with the development and commercial availability of the dc-Superconducting Quantum Interference Device (dc-SQUID), a very sensitive flux-to-voltage converter, that allows to easily match the TES low impedance.

The change of the TES resistance when heated by the incident radiation is read by a dc SQUID as is shown in Fig. 2. A change of the current in the integrated input coil $L_i$ of the SQUID is converted via the mutual inductance $M_i$ in a change of magnetic flux through the SQUID loop resulting in a voltage output signal.

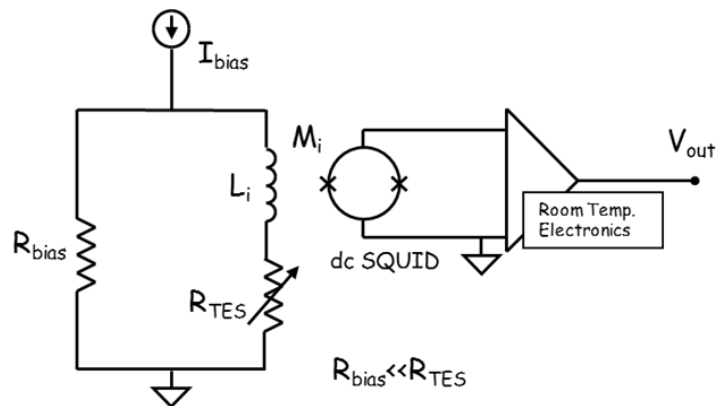

**Fig. 2** An example of a SQUID readout circuit of a voltage biased TES.

The Joule heating in a current-biased TES can lead the detector into the normal state (positive electrothermal feedback) making impossible a stable operation in the narrow superconducting transition region. This problem is solved by voltage-biasing the TES (negative electrothermal feedback, see Fig. 2). The voltage, determined by the resistance $R_{bias} \ll R_{TES}$, is chosen to put the TES in its so-called "self-biased region" where the power dissipated in the device is constant. When the incident radiation is absorbed, the extra power increases the TES resistance, causing a drop in the TES current; then the Joule power drops, allowing the device to cool back to its equilibrium state in the self-biased region. As the TES is operated in series with the input coil of the SQUID, a change in the TES current produces a change in the flux to the SQUID loop, whose output is amplified and read by room-temperature electronics.

TESs were developed for measurement of single photons from gamma and X-ray [3] down to the visible and near infrared region [4, 5]

where they demonstrated photon-number-resolving capability but better sensitivities are possible.

The TES-calorimeter sensitivity is basically limited by the magnitude of the thermal energy fluctuations but when it is voltage-biased operated with negative electrothermal feedback and the optimal filter to an exponential pulse is applied, its energy resolution $\Delta E$ can be shown to be $\approx 2.3(4k_B T_c^2 C_e(1/\alpha))^{1/2}$ [6], where $k_B$ is the Boltzmann constant, $T_c$ the critical temperature of the TES superconductor and $C_e$ its heat capacitance. In addition, $\alpha = (T/R)(dR/dT)$ is a unitless measure of the sharpness of the superconducting transition and $C_e = \gamma V T_e$, where V is the sensor volume, $T_e$ the electron temperature, and $\gamma$ the Sommerfeld coefficient.

The above $\Delta E$ expressions suggest lowering the operation temperature and the sensor volume as much as possible in order to detect the low-energy microwave photons. For instance, by using TiAu bilayer as TES materials ($V_{Ti}$ = 300 × 80 × 10 nm$^3$, $V_{Au}$ = 300 × 80 × 25 nm$^3$, $\gamma_{Ti}$ = 3.35 mJ/K$^2$/mol, $\gamma_{Au}$ = 0.69 mJ/K$^2$/mol) it is necessary to reach a critical temperature of 40 mK and an $\alpha$ of 20 to expect an energy resolution $\Delta E \approx$ 0.1 meV which corresponds to a frequency resolution of 24 GHz.

In order to finely tune the transition temperature in the range of tens of mK, we are investigating two different approaches: the proximity effect in a normal and superconducting bilayer (Fig. 3 *Left*) and the injection of an extra current in the TES film (Fig. 3 *Right*).

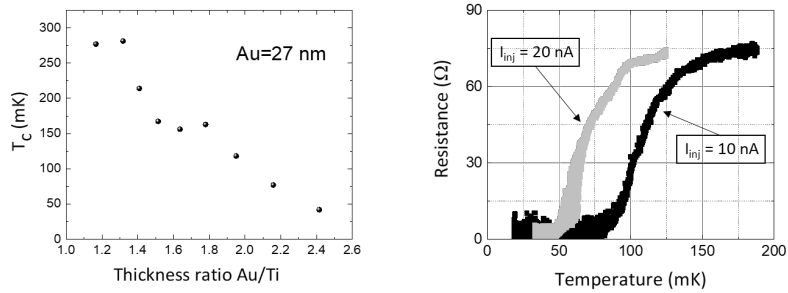

Fig. 3 *Left* Transition temperature of a AuTi bilayer film as a function of the thickness ratio Au/Ti. *Right* Tuning of the transition temperature by injection of an extra current in an Al(10 nm)Cu(15 nm) bilayer TES.

## 3 Current biased Josephson Junction

Superconducting circuits based on Josephson junctions [7] are fabricated by standard lithographic technologies and can be integrated in complex systems, with an extremely high flexibility and controllability. The

# Title

nonlinearity of the Josephson junctions allows the realization of systems behaving as artificial atoms, with typical level spacing of the order of few GHz up to tens of GHz. All these characteristics make them interesting as detectors for microwave single photons. Non-destructive measurement of 6 GHz photons was performed in a coplanar resonator [8]. Further improvements were obtained using two-cavity schemes and 3D cavities [9-11]. The flexibility of Josephson elements makes possible a series of different approaches. For example, it is possible to consider a current biased Josephson junction operating in the phase regime, where the escape from the excited level causes a transition to the voltage state that can be easily read out [12,13], or a flux qubit based on a lambda-type three level system [14].

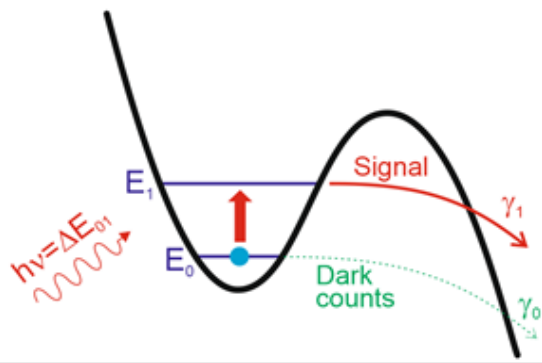

**Fig. 2** Washboard potential describing the dynamics of the superconducting phase of a CBJJ.

Our aim is to drastically reduce the dark count of this type of devices below the Hz rate (the possibility to go below the mHz will be also explored). Our approach to achieve this goal can be summarized in 3 steps:

1. Aluminum junctions are fabricated and current-biased in order to get plasma frequency $\nu_p$ =14 GHz and a potential barrier height about $2\nu_p$. With this configuration and operating at 20 mK, transitions to the junction resistive state from the ground state are at a rate of the order of 15 kHz. Note that, for an axion experiment working at this frequency, this dark count rate is equivalent to work with a quantum-limited linear amplifier. Single photons will induce transitions from ground state to the first excited state with high efficiency. Tunneling transition-rate from the excited state are expected to be about 15 MHz.

2. To improve the switching efficiency we can increase the tunneling rate by sending microwave power of frequency $(E_2-E_1)/h$ to the junction, exploiting the system non-linearity, to further excite the junction from the first to the second excited-level. Tunneling rate from the second excited level is expected to be a factor 1000 larger, about tens of GHz, enough to cope with junction dissipation. Dark count rate can then be further lowered increasing the potential-barrier height suitably varying the bias current.

    3. The possibility to excite an overtone transition from ground state to the fourth excited state will be investigated. These transitions, forbidden for harmonic oscillator, are allowed in anharmonic ones. For instance, setting the plasma frequency 3.5 GHz and the potential barrier to 14GHz, the tunneling transition rate from the ground state is suppressed down to mHz level. A single photon with frequency $(E_4-E_0)/h$ (about 14 GHz) is still expected to be absorbed with good efficiency, triggering the tunneling transition from the higher excited state. A different possibility are overcurrent induced transitions as proposed in [15].

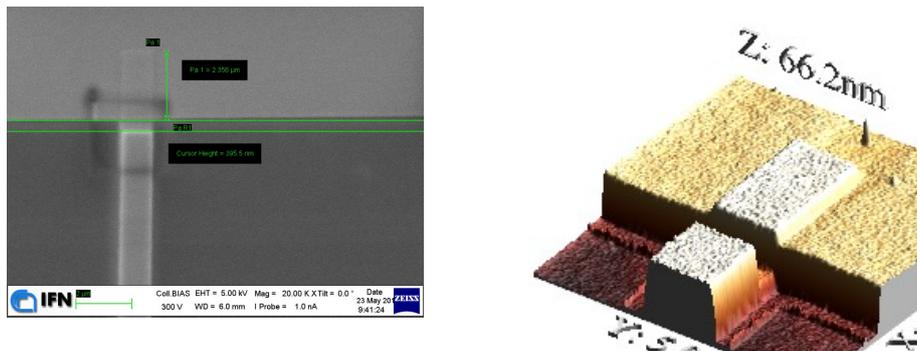

**Fig. 4** *Left* First test sample of aluminum/aluminum oxide/aluminum junction fabricated by Electron Beam Lithography, shadow mask evaporation, in situ oxidation and lift off. *Right* Atomic Force Microscopy scan of the junction.

## 4 Conclusions

We have presented the project Simp for the development of a single microwave photon detector for fundamental physics experiments, its goals and the planned techniques to achieve them. Some preliminary results have been also presented.

# Title